\documentclass[useAMS,usegraphicx,usenatbib]{mn2e}
\usepackage[a4paper,centering, totalwidth=520pt, totalheight=700pt]{geometry}
\usepackage{graphicx}
\usepackage{aas_macros}     
\usepackage{amsmath}
\usepackage{amssymb}
\usepackage{fixltx2e}[1999/12/01]

\newcommand{\Mass}{ \ensuremath{ h^{-1} M_{\odot}} }

\newcommand{\Mpc}{ \ensuremath{h^{-1} {\rm Mpc}} }

\newcommand{\vect}[1]{\bmath{#1}}

\title[Extending the halo mass resolution of $N$-body simulations]
{Extending the halo mass resolution of $N$-body simulations}
\begin{document}
\setlength{\topmargin}{-1.cm}

\author[Angulo et al.]{
\parbox[h]{\textwidth}
{Raul E. Angulo$^{1,2,3} \thanks{rangulo@cefca.es}$, 
Carlton M. Baugh$^{3}$, 
Carlos S. Frenk$^{3}$,
Cedric G. Lacey$^{3}$.}
\vspace*{6pt} \\
$^{1}$Centro de Estudios de F\'isica del Cosmos de Arag\'on, Plaza San Juan 1,  Planta-2, 44001, Teruel, Spaini.\\
$^2$  Kavli Institute for Particle Astrophysics and Cosmology, Stanford University, SLAC National Accelerator Laboratory, \\
Menlo Park, CA 94025, USA \\
$^3$ Institute for Computational Cosmology, Dept. of Physics, Durham University, South Road, Durham DH1 3LE, UK.}
\maketitle

\date{\today}
\pagerange{\pageref{firstpage}--\pageref{lastpage}} \pubyear{2013}
\label{firstpage}

\begin{abstract} 
We present a scheme to extend the halo mass resolution of 
$N$-body simulations of the hierarchical clustering of dark 
matter. The method uses the density field of the simulation 
to predict the number of sub-resolution dark matter haloes 
expected in different regions. The technique requires as 
input the abundance of haloes of a given mass and their 
average clustering, as expressed through the linear and higher order 
bias factors. These quantities can be computed analytically or, 
more accurately, derived from a higher resolution simulation as 
done here. Our method can recover the abundance and clustering  
in real- and redshift-space of haloes with mass below $\sim 7.5  
\times 10^{13}h^{-1}M_{\odot}$ at $z=0$ to better than $10\%$. 
We demonstrate the technique by applying it to an ensemble of 50 
low resolution, large-volume $N$-body simulations to compute the 
correlation function and covariance matrix of luminous red galaxies 
(LRGs). The limited resolution of the original simulations results 
in them resolving just two thirds of the LRG population. 
We extend the resolution of the simulations by a factor of 30 in halo 
mass in order to recover all LRGs. 
With existing simulations it is possible to generate a halo catalogue 
equivalent to that which would be obtained from a $N$-body simulation 
using more than 20 trillion particles; a direct simulation 
of this size is likely to remain unachievable for many years.  
Using our method it is now feasible to build the large numbers 
of high-resolution large volume mock galaxy catalogues required to 
compute the covariance matrices necessary to analyse upcoming galaxy 
surveys designed to probe dark energy.  
\end{abstract}

\section{Introduction}

The spatial distribution of galaxies is an important resource 
in physical cosmology, encoding information about the physics 
of galaxy formation and the values of the basic cosmological parameters 
\citep{Guzzo2008,Cabre2009,Sanchez09,Beutler2011,Zehavi2011,
Sanchez2012,Reid2012}. 
A number of galaxy surveys are underway or planned which share the 
primary science goal of using the large-scale structure of the 
Universe to constrain the nature of dark energy 
\citep[e.g.][]{Schlegel2011, Laureijs2011}. 
To achieve this, 
these surveys will map galaxies over many tens of cubic gigaparsecs. 
As the clustering signals predicted by competing cosmological models 
are often very similar, the scientific exploitation of the surveys 
will be limited by how well we are able to understand the systematic 
errors which may affect statistical measures of the large-scale structure of 
the Universe. 

A complete understanding of the systematic and sampling errors 
associated with clustering measurements requires many effects to be 
modelled, including cosmic variance, nonlinear evolution of density 
fluctuations, scale-dependent bias, redshift space distortions, 
discreteness effects and survey geometry. To meet the challenge of 
providing the best possible theoretical predictions, the most accurate 
techniques have to be employed. Currently this means using 
$N$-body simulations of the hierarchical clustering of the dark matter (DM) 
\citep[see][]{Springel2006}. 
The need to model clustering accurately on scales beyond 
$100 h^{-1}$Mpc requires computational boxes in excess 
of $1 h^{-1}$Gpc on a side \citep{Angulo2008}. 
Resolving Milky-Way mass haloes or smaller in such calculations is expensive 
but has been achieved in a small number of cases 
\citep[for a summary of the state of the art, see ][]{Kuhlen2012}. 
Such calculations are currently one-offs and the computational 
resources are not available to generate the large numbers of such runs 
which are required to compute covariance matrices for large-scale 
structure statistics.  

The principal way to study errors on clustering measurements from 
galaxy surveys is through an accurate model of the experiment itself 
\citep{Baugh2008}. For the case of relevance here (the spatial 
distribution of galaxies), this is optimally achieved in a three 
step process. First, the halo clustering is predicted by following 
the evolution of particles in a $N$-body simulation 
\citep[see the recent reviews of][]{Springel2006,Kuhlen2012}. 
Second, the properties of galaxies within these haloes are 
predicted using a semi-analytical 
model of galaxy formation \citep[for reviews see][]{Baugh2006,Benson2010}. 
And finally, the appropriate flux limit, sample selection, redshift 
completeness and the geometry of the survey need to be applied to the 
catalogues \citep[e.g.][]{Merson2013}. Some of these steps may be 
modified. For example, in the case of a low resolution simulation, 
``galaxies" may be added using an empirical rule based on the 
smoothed density of the dark matter 
\citep{White1987,Cole1998}. 
The predictions of the semi-analytical model may be substituted 
by empirical techniques tuned to match observational data, such 
as halo occupation distribution modelling or sub-halo abundance 
matching \citep{Zehavi2011,Simha2012}. 

The direct approach to modelling the errors on clustering statistics, 
namely populating many large volume, high resolution $N$-body simulations 
with galaxies using semi-analytical galaxy formation models, 
is computationally expensive for two reasons: 
i) The large number of independent $N$-body simulations 
needed to make robust estimates of the errors. An adequate estimate of 
the variance requires several dozen realizations of the density field 
(e.g. a $10$\% error on the variance for a Gaussian distribution requires 
$\sim 50$ realizations). An order of magnitude more simulations is needed 
to robustly compute the full covariance matrix \citep{Takahashi2009}. 
ii) The huge dynamic range required to resolve the haloes which are likely 
to host the galaxies observed. For instance, in $N$-body simulations with box sizes of 
a few gigaparsecs, only Milky-Way sized haloes can be identified robustly, 
even in the highest mass resolution simulation of this type carried out to date 
\citep{Angulo2012a}. The mass limit is much larger in typical 
large-volume simulations \citep{Fosalba2008, Teyssier2009, Kim2009, Alimi2012}. 
In fact, currently there is no simulation which can simultaneously model, 
for instance, the faintest galaxies and the volume to be probed by the 
Dark Energy Survey \citep{DES2005}. Although algorithms and computer hardware 
are constantly improving, finite computational resources impose a limit on
$N$-body simulations: carrying out a single simulation, not to
mention an ensemble of them, meeting the desired requirements is 
currently prohibitively expensive computationally.

Several authors have proposed algorithms to predict galaxy clustering 
efficiently and to overcome the difficulties stated above. Amongst 
the simplest are realizations of Gaussian or log-normal density fields 
\citep[e.g.][]{Mesinger2007,Percival2001}. More sophisticated ideas have 
been implemented using second order perturbation theory \citep{Monaco2002, 
Scoccimarro2002, Kitaura2013, Kitaura2013b, Manera2013, Monaco2013}. 
In a different approach, the use of simulation particles to mimic galaxy 
clustering has been adopted in several studies by invoking a 
prescription based on the local DM density \citep{Cole1998, Cabre2008}. 
\cite{White2013} recently proposed an extension to these ideas, using 
low resolution particle-mesh $N$-body simulations to generate large numbers 
of realisations of the DM density field, which is sampled to mimic the 
clustering of different samples of DM halos. However, none of these 
approaches has fully achieved the combination of simplicity and 
accuracy desirable when modelling a given cosmological experiment.

The objective of this paper is to present and test a scheme
to create mock catalogues of the large-scale distribution of galaxies 
in a computationally inexpensive way.\footnote{The method was introduced in the PhD 
Thesis of the lead author \citep{AnguloPhD}.} Our approach uses 
the DM density field extracted from $N$-body simulations to predict a 
halo population whose properties can be derived from a higher resolution 
simulation or analytically. 
This effectively extends the halo mass resolution of $N$-body simulations
down to an arbitrarily low limit.
Note, a very similar method has been developed independently by 
\cite{delaTorre2013}. Here, we present an alternative formulation 
of the method together with an enhanced suite of tests, which focus 
on larger scales, specifically the creation of covariance matrices 
for the two-point correlation function on the baryonic acoustic 
oscillation (BAO) scale. 

The structure of this paper is as follows. 
In \S \ref{b:sec:method} we provide details of
our method along with its theoretical motivation. In \S \ref{b:sec:results} we
apply our algorithm to an ensemble of $N$-body simulations to investigate the
limitations and range of applicability of the procedure. 
The haloes created by our algorithm can be combined with higher mass haloes 
which are identified directly in the
the simulations to extend the range of halo masses in the 
simulation box. The resulting hybrid halo catalogue can be fed into a
semi-analytic galaxy formation model or combined with a HOD model. 
To illustrate the feasibility of the idea, our procedure is shown in
action in \S \ref{b:sec:lrg} where we use a HOD to predict the errors on the
clustering of luminous red galaxies (LRGs). Finally, in \S \ref{b:sec:conc}, 
we present a summary and discussion of our findings.

\section{Method} \label{b:sec:method}

In this section we present the algorithm used to generate a 
halo population from the density field in DM simulations. We start by giving the
motivation and main ideas behind the method (\S \ref{b:sec:method:mot}) 
and then we outline the steps to be followed in a practical 
implementation of the technique (\S \ref{b:sec:method:app}). 

\subsection{Theoretical Motivation} \label{b:sec:method:mot}

Assuming that the abundance of haloes at a given position,  
${\bf x}$, is a function of the local underlying nonlinear 
DM density alone, we can write the number density field 
of haloes of mass, $M$, as:

\begin{equation}
\delta_{\rm h, R}({\bf x}, M) = f_M(\delta_{\rm dm, R}({\bf x})),
\label{b:eq:loc}
\end{equation} 

\noindent where $f_M$ is a smooth and arbitrary function (which could, 
in principle, be different for haloes of different mass), $\delta({\bf x})$ 
is the density contrast, defined as $\rho({\bf x})/\langle \rho({\bf x}) \rangle -1$, 
where $\rho({\bf x})$ is the density at ${\bf x}$ and $\langle \rho({\bf x}) \rangle$ 
is the mean density,  
and the subscripts $\rm h$ and $\rm dm$ refer to the density field of 
haloes and dark matter respectively. $R$ is the scale on which both 
density fields are smoothed and is set by the smallest scale on 
which Eq.~\ref{b:eq:loc} holds. 

On sufficiently large scales the DM density approaches the mean value, 
$|\delta_{\rm dm}(x|R)| \ll 1$, which allows us to express 
Eq.~\ref{b:eq:loc} as a Taylor series expansion in 
$\delta_{\rm dm}$ \citep[see e.g][]{Fry1993}:

\begin{equation} 
\delta_{\rm h,R}(x,M) = \sum_{k=0}^{\infty} \frac{b_k(M)}{k!} \delta_{\rm dm, R}^{k}(x), 
\label{b:eq:dhalo}
\end{equation}

\noindent where the subscript $R$ denotes the smoothing scale. 
The coefficients $b_{k}$ are usually referred to as the
bias parameters. In particular $b_{1}$ is known as the linear bias. 
These parameters can be derived analytically from collapse models 
\citep{Mo1997} or measured directly from $N$-body simulations 
\citep{Angulo2008b}. Note that the functional form adopted in 
Eq.~\ref{b:eq:dhalo} is not the only possibility; for instance, 
\cite{delaTorre2013} invoke an exponential model. However, both models
converge asymptotically on large scales.

It is straightforward to write down an expression for 
the expected number density of haloes of a given mass 
in a region in which the DM density field has been 
smoothed, 

\begin{eqnarray} 
N_{\rm h, R}({\bf x},M) = \langle N_{\rm h, R}({\bf x}, M) \rangle \times \nonumber \\
                 \left[ b_0+ b_{1} \delta_{\rm dm, R}({\bf x}) + 
                  \frac{b_2}{2} \delta_{\rm dm, R}^{2}({\bf x}) + 
                  \mathcal{O}(\delta_{\rm dm, R}^{3}) \right].
\label{b:eq:nhalo}
\end{eqnarray}

\noindent Here the brackets $\langle \rangle$ denote an average over 
all smoothing regions, and so $\langle N_{\rm h, R}({\bf x}, M) \rangle$ 
is the standard halo mass function. Note that $b_0$ is set by
requiring that the expresion inside square bracktes is equal to 
the unity when averaged over all regions.
As we discuss below, it is possible to use this expression to 
construct a halo density field which displays the halo 
abundance and clustering properties expected in a $N$-body simulation.

\subsection{Implementation} \label{b:sec:method:app}

It can be seen clearly that, under our assumptions, the expected abundance of
haloes at a given location (Eq.~\ref{b:eq:nhalo}), depends on three quantities:
i) the dark matter density field at the location, ii) the mean number density
of haloes of a given mass and iii) the bias parameters as a function of halo
mass.  The core of our method is that it is possible to recover the underlying
DM density field directly from simulations with high fidelity (even in the case
of low-resolution simulations), and also that both the bias parameters and the
mean number of haloes can be calculated easily, either analytically or from
high-resolution $N$-body simulations (which will typically be of much smaller
volume than the simulations we wish to populate with haloes). As a consequence
of bringing these ingredients together, a population of DM haloes, which spans
an {\it arbitrarily wide range of masses}, can be created.

Subject to the validity of our assumptions as discussed below, the population
of haloes generated using our method has, by construction, the correct
abundance and clustering on scales larger than the chosen smoothing scale. In
fact, not only are the two-point statistics   reproduced for the halo
distribution but, in principle, the correct volume-averaged higher order
statistics are also recovered  (as can be seen from Eq.~\ref{b:eq:dhalo}). We
call the haloes generated using our technique ``sub-resolution" haloes. In the
next section we will test our method by applying it to generate all of the
haloes in a simulation volume, in order to assess the validity of the approach.
In practice we will use a hybrid halo catalogue made up of haloes which are
resolved directly in the simulation, and lower mass haloes which are 
added using our technique, hence the name ``sub-resolution". 

There are, inevitably, limitations in the sub-resolution halo catalogues which
arise from our simplified treatment of halo formation. First, our expressions
are only strictly valid when the density contrast is small, $\delta_{\rm dm}
\ll 1$.  This sets a minimum smoothing scale that can be used which in turn
determines the smallest scale on which the halo clustering can be reproduced.
Second, in a practical implementation, Eqs.~\ref{b:eq:dhalo} and
~\ref{b:eq:nhalo} have to be truncated at a given order which creates two
problems: i) The clustering statistics of orders higher than the truncation can
not be reproduced accurately.  
There will be some information about the higher 
order clustering of haloes since we are applying our technique to 
the evolved density field in the DM simulation. 
ii) In underdense regions Eq.~\ref{b:eq:nhalo} can predict
a negative number of massive haloes.  This would happen in an expansion
truncated at first order if $b(M) > 1$ and $\delta_{\rm dm} < - 1/b$, implying
$\delta_{\rm hh} < -1$.  Consequently, we expect our procedure to break down
for haloes more massive than $M_{*}$. These restrictions are not prohibitive
though, since our algorithm is primarily designed to add low-mass and therefore low-bias halos.
Moreover, as discussed by \cite{delaTorre2013}, the small-scale clustering in
magnitude limited galaxy samples tends to be dominated by satellite galaxies
hosted by massive halos. Since these are typically resolved directly in
$N$-body simulations, a relatively large smoothing scale does not introduce
noticeable artifacts even in the small-scale clustering of catalogue
constructed using our algorithm. We investigate and quantify these restrictions
in the following sections where we present our algorithm in action.

\section{Testing the method} \label{b:sec:results}

We now apply and test the procedure outlined in the previous section.  In
\ref{b:sec:lb} we provide details of the implementation of the method and
present some general characteristics of the resulting halo catalogues. In
section \ref{b:sec:test} we show the results of three basic tests and a
comparison with haloes identified directly in a high resolution $N$-body
simulation. The sub-resolution catalogues we generate in this section cover a
wide range of halo masses, including those of haloes that are resolved in the
$N$-body simulations.  The goal in this section is to establish the range of
validity of our method in view of the assumptions and approximations which
underpin it. As we pointed out in the previous section, the actual
implementation of the method (\S~4) will make use of ``hybrid" halo catalogues
in which the higher mass haloes are those directly resolved in the simulation
and the lower mass ones are the ``sub-resolution'' population generated by our
algorithm.  

\subsection{The sub-resolution halo catalogue} \label{b:sec:lb}

To characterise the performance of our method we use the simulations described
in \cite{Angulo2008}. These include a suite of $50$ low-resolution simulations,
referred to as the {\tt L-BASICC} ensemble. Each of these modelled the
gravitational interactions between $448^3$ particles of mass
$1.85\times10^{12}\,h^{-1}{\rm M_{\odot}}$ in a periodic box of side
$1340\,h^{-1}M_{\odot}$.  We also employ a higher-resolution run, dubbed {\tt
BASICC}, which used $1448^3$ particles of mass $5.49\times10^{10}\,h^{-1}{\rm
M_{\odot}}$, also in a periodic box of side $1340\,h^{-1}M_{\odot}$.  Note that
one of the {\tt L-BASICC} simulations has exactly the same initial density
field as used in the {\tt BASICC} run.  Haloes are identified in the simulation
outputs using a Friends-of-friends (FoF) percolation algorithm
\citep{Davis1985}. We stress that it is computationally inexpensive to
carry out such a set of low resolution simulations. Each of the {\tt L-BASICC}
would only take approximately 150 CPU-hours on modern supercomputers.

\begin{figure*}
\includegraphics[width=18cm]{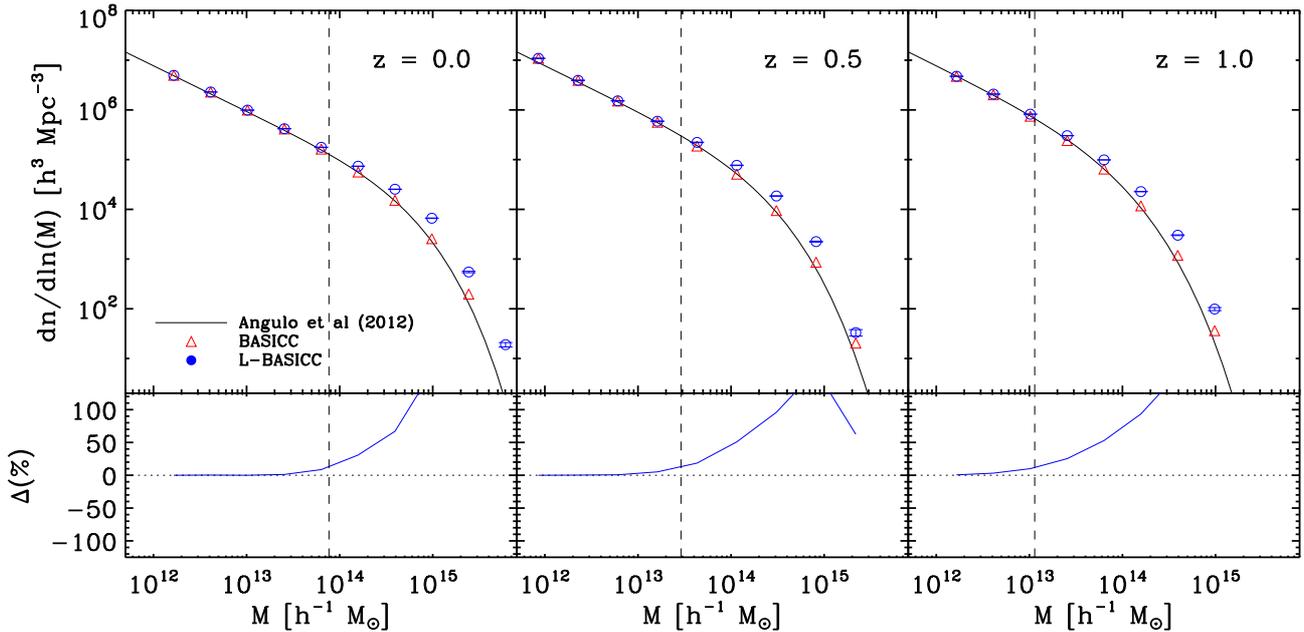}
\caption[The mean number of haloes per logarithmic mass bin, as a function of
their mass.]{
A test of the method showing the comparison 
between the mass function of FoF haloes resolved
in the  {\tt BASICC} simulation (red triangles) and the mean of 
sub-resolution halo catalogues built from the {\tt L-BASICC} ensemble 
(blue circles). Note that in this test case sub-resolution haloes are 
generated across the whole mass range plotted to assess the range of 
validity of the technique. 
Top: The mean number of haloes per logarithmic mass bin, as a 
function  of their mass. The error bars show 
the dispersion from applying our algorithm to 50 simulations ({\tt L-BASICC}). 
Each column shows a different redshift and the empirical fit to various 
$N$-body results from \cite{Angulo2012a} (solid lines). The 
vertical dashed lines indicate the halo mass at which the number of haloes 
resolved directly in the simulations and those created by our algorithm first 
differ by $10\%$ moving in the direction of increasing mass. 
The relative difference between these catalogues is shown by the curves 
in the bottom panels.
\label{b:fig:mf}}
\end{figure*}

Following the algorithm described in \S~\ref{b:sec:method} we computed a
sub-resolution halo catalogue for the three outputs ($z=0$, $0.5$ and $1$) of
each of the $50$ simulations in the {\tt L-BASICC} ensemble. 
This process is made up of three steps. The first is the construction 
of the DM density field in the simulations. This is performed by placing 
particles onto a grid using the nearest grid point mass assignment scheme 
\citep{Hockney1981}. We use a grid of $256^3$ cells 
(the cell size is $5.2\,\Mpc$) which is set so 
that $\langle\delta^2\rangle \sim 1$. We therefore expect to obtain 
an inaccurate estimation of the halo clustering on scales smaller 
than a few times the size of the grid cell. Note that \cite{delaTorre2013} 
followed an alternative path and constructed the DM density field field 
from the resolved halo population. As there are fewer halos than
particles, there is a larger amount of noise in the reconstructued density
field. Dealing directly with simulation particles also allows the use of
Lagrangian smoothing techniques \citep{Abel2012, Shandarin:2012}, which have
been recently shown to have extremely low discreteness noise
\citep{Angulo2013b,Angulo2013c}.

The next step is to tabulate the halo bias parameters and the number density of
haloes as a function of mass. We extract these relationships from the 
higher resolution {\tt BASICC} simulation in logarithmic mass bins of
width $\Delta \log_{10} M = 0.426$. Both quantities are computed by smoothing
the haloes and DM field in $256^3$ cells and then averaging the 
values across the grid. 

Finally, these three quantities are brought together to compute the 
expectation value for the number density of haloes on every point of 
the grid. There are several points regarding the placement of haloes that are
worth noting. i) The actual number of haloes in each cell is generated from a
Poisson distribution with the expectation value as the mean. In doing this we
have also neglected the covariance between halo mass bins, which is justified
given the box size of our simulations \citep{SmithMarian2011}. ii) The haloes
are placed randomly within each of the smoothing volumes. iii) Each of these
haloes is given a peculiar velocity equal to the mean velocity of the DM
particles within the same cell. Alternatively one could use some sort of
interpolation scheme such as that used by \cite{delaTorre2013}.  iii)
Eq.~\ref{b:eq:nhalo} is truncated at linear order. 

As a result of following this procedure we obtained $50$ independent
sub-resolution halo catalogues at the three redshifts mentioned above.  Each
contains approximately $17$ million haloes with mass between $5.48\times
10^{11}\,\Mass$  and $1\times 10^{16}\,\Mass$ at $z=0$.  In the following
subsection we will explore the properties of these catalogues.

\subsection{Abundance and clustering} \label{b:sec:test}
\begin{figure*} 
\includegraphics[width=18cm]{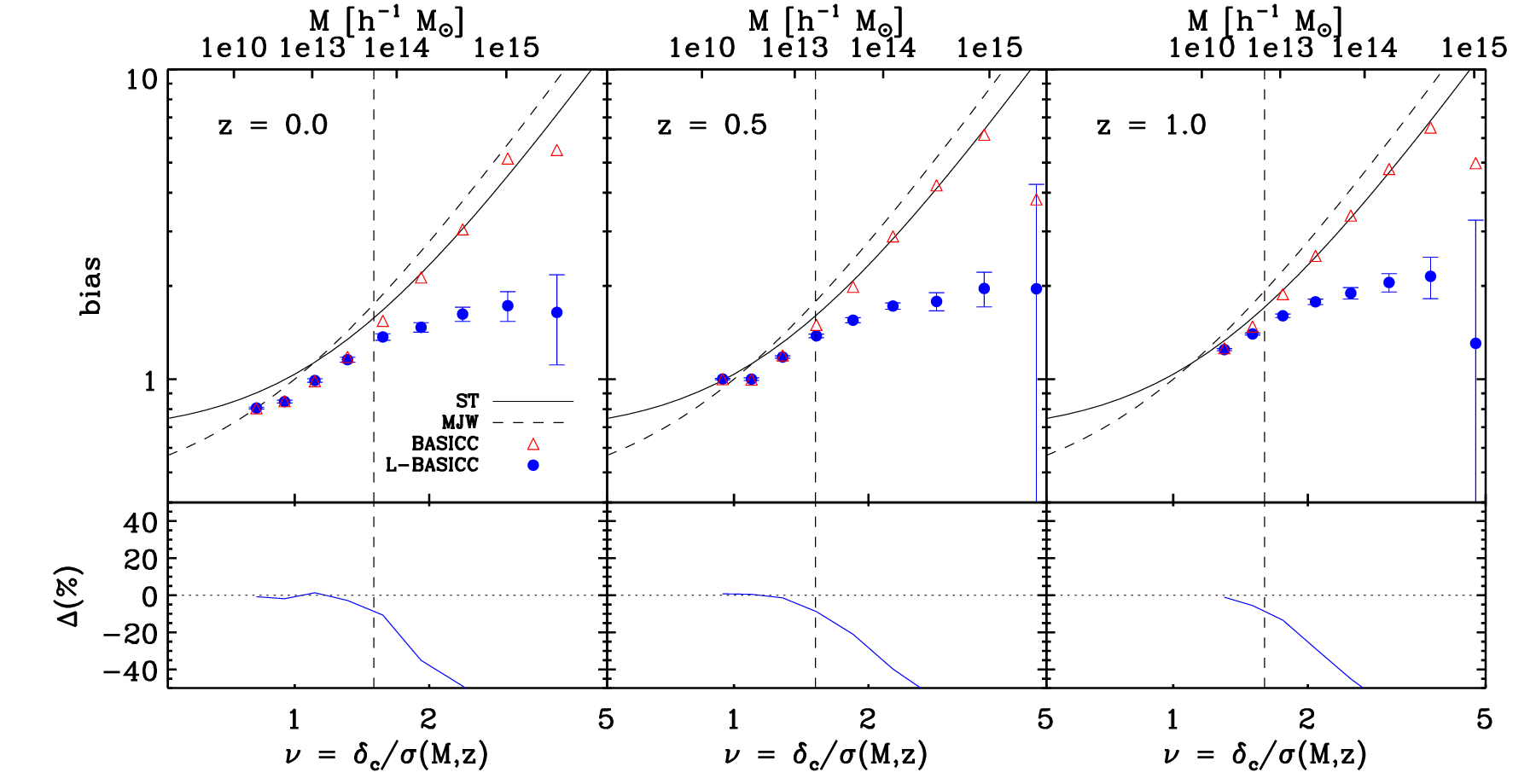}
\caption[The linear bias parameter as a function of halo mass.]{
A test of the clustering predicted by our method. In this test 
case sub-resolution haloes are generated for the full mass range 
plotted to assess the range of validity of the technique. Top: The linear
bias parameter as a function of halo mass (top axis) and peak height 
$\delta_{\rm c}/\sigma(M,z)$ (bottom axis). Each panel shows a different 
redshift as labelled. The blue filled circles with error bars show the 
mean bias and the dispersion for the sub-resolution haloes generated by 
our algorithm applied to the 50 low resolution {\tt L-BASICC} simulations. 
The bias measured from FoF haloes in the high-resolution {\tt BASICC} 
simulation is shown as red triangles.
Theoretical predictions from \cite{Mo1997} and \cite{Sheth2001}
are also shown using different line styles as labelled. 
The lower panels show the relative difference between the halo bias 
from the two different catalogues.
The vertical dashed lines show where the clustering of haloes in 
the sub-resolution catalogue first differs from that measured from 
the FoF haloes by 10\%, moving in the direction of increasing mass. 
\label{b:fig:bias}}  
\end{figure*}

In this subsection we compare the abundance and clustering strength in our
sub-resolution halo catalogues with the same quantities measured using haloes
directly identified by a FoF algorithm in a high resolution simulation
\citep[for details of the FoF catalogues see][]{Angulo2008}.

The upper panels of Fig.~\ref{b:fig:mf} show the differential halo mass
function from our catalogues (blue filled circles) and that from FoF haloes
identified in the {\tt BASICC} simulation. In the lower panels we can see the
differences between the two populations more clearly on a linear scale.  This
figure shows that there is excellent agreement between the number of haloes
generated using our algorithm and that obtained directly in the higher
resolution $N$-body simulation. This represents an initial validation of the
ideas and their implementation presented in this paper.  Our method predicts an
abundance of haloes that agrees with the direct simulation results to better
than $10\%$ for objects of mass $M < 7.51 \times 10^{13}\,\Mass$ at $z=0$, $M <
2.7 \times 10^{13}\,\Mass$ at $z=0.5$ and $M < 1.14 \times 10^{13}\,\Mass$ at
$z=1$. There is a strong disagreement between the numbers of sub-resolution and
FoF haloes at the high mass end. This is caused by the fact that
Eq.~\ref{b:eq:dhalo} is inconsistent for highly biased haloes in low density
regions where $\delta_{\rm h} < -1$ (the problem  is alleviated in the low mass
regime where $b \lesssim 1$) and because we have truncated Eq.~3 at the linear
bias term. As a consequence of haloes of a fixed mass becoming more biased with
increasing redshift, the mass function of sub-resolution haloes provides an
acceptable match to the simulation results (i.e. better than 10\% agreement)
over a reduced range of masses at high redshift. 

We extend the comparison by investigating the clustering strength in the
sub-resolution catalogues. Each column of Fig.~\ref{b:fig:bias} displays the
linear bias parameter as a function of the peak height\footnote{Here $\delta_c$ is the threshold for collapse
in linear perturbation theory and $\sigma(M,z)$ is the linear theory {\it rms}
variance in the density field smoothed on a scale enclosing mass $M$ at
redshift $z$.} 
,
$\delta_c/\sigma(M,z)$
on the bottom axis and as a function of mass on the top axis.
Note that we compute the linear bias, $b$, by smoothing the halo and DM density
fields in cells of size $167\,\Mpc$, and taking the ratio i.e. $b = \langle
\delta_{\rm hh}\rangle / \langle \delta_{\rm mm} \rangle$.  As in the previous
plot, the vertical lines indicate the maximum halo mass at which the result
from the sub-resolution haloes agrees to within $10\%$ with that of the
resolved haloes. Similar to the behaviour seen in Fig.~\ref{b:fig:mf}, at the
high mass end, the sub-resolution haloes fail to reproduce the clustering
measured from the resolved FoF catalogues, which suggests a common origin for
the discrepancies seen in the abundance and clustering of sub-resolution haloes
at high masses.  Note, that the $10\%$-difference mass limit derived from the
clustering comparison is slightly smaller than that derived from the mass
function at; $z=0$ $M_{\rm max} = 5.23\times 10^{13}\,\Mass$ while at $z=0.5$
and $z=1$ $4.3 \times 10^{13}\,\Mass$ and $6.73 \times 10^{12}\,\Mass$
respectively. Again, the very good agreement apparent at low masses validates
our approach. 

\begin{figure*}
\includegraphics[width=18cm]{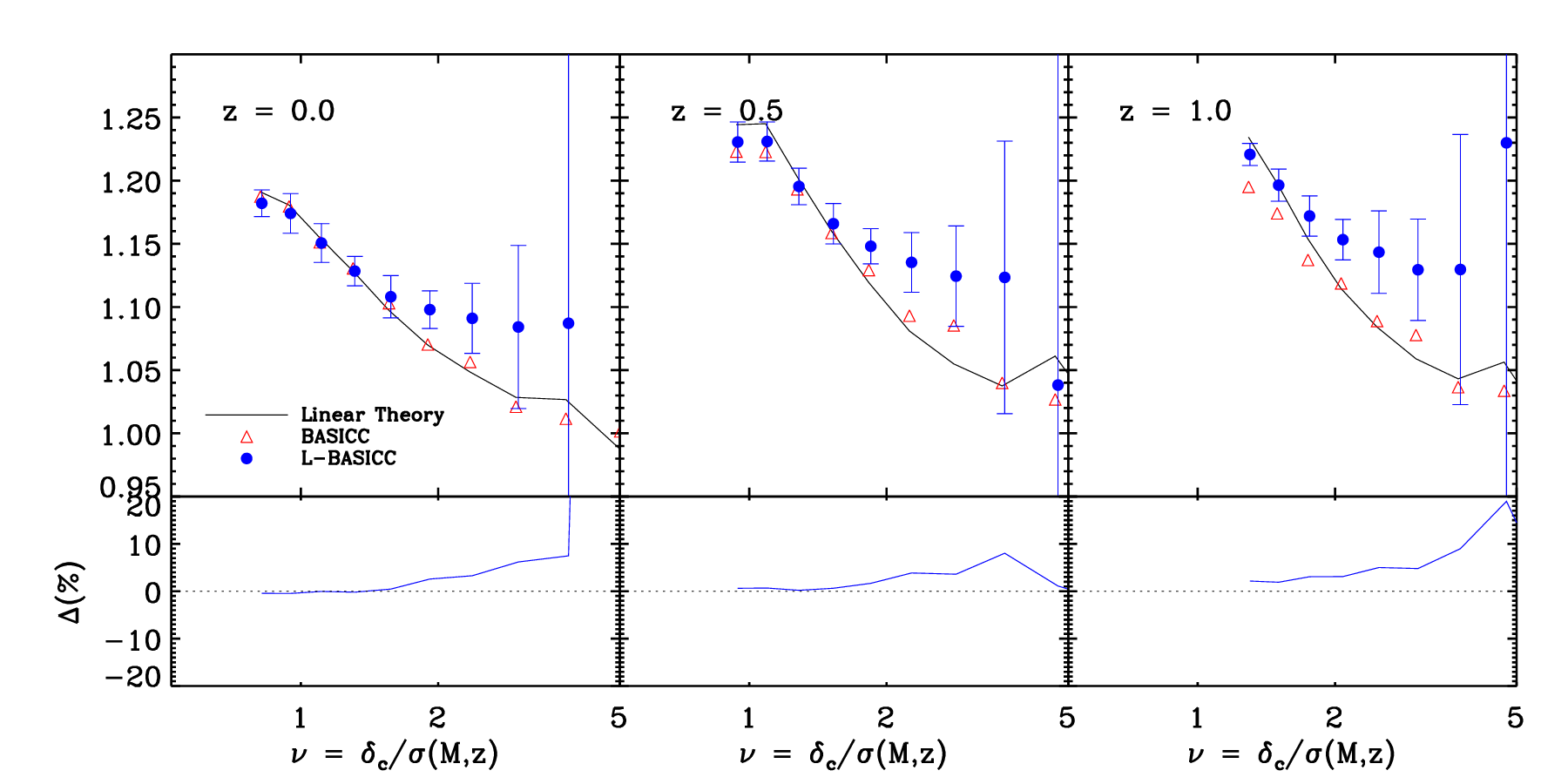}
\caption[The linear bias parameter for sub-resolution haloes measured in
redshift space divided by that measured in real space.]{
A test of the peformance of the model predictions for the clustering 
measured in redshift-space. In the test case sub-resolution haloes are 
generated across the full range of halo masses plotted to evaluate the performance of the technique. The linear bias
parameter for haloes measured in redshift space, $b_{\rm z}$, divided by 
that measured in real space, $b_{\rm r}$, as a function of the peak height  
(bottom axis). The mean and dispersion of this 
quantity, measured from our sub-resolution haloes in an ensemble of 
low resolution simulations, is displayed using blue symbols with error bars.  
We display the results measured from FoF haloes in the {\tt BASICC} 
simulation using red triangles. For comparison, we have also
included the prediction based on linear theory 
(solid line, see Eq.~\ref{b:eq:kaiser}). The lower panels show 
the relative difference between the sub-resolution results and those 
obtained from the {\tt BASICC} run. 
\label{b:fig:biasz}}
\end{figure*}

Finally, we explore the spherically averaged clustering of the halo catalogues
in redshift space. Fig.~\ref{b:fig:biasz} shows the ratio between the linear
bias parameter measured in redshift space and that measured in real space for
the sub-resolution haloes and for the FoF haloes. In linear perturbation
theory, this quantity is equivalent to the square root of the Kaiser ``boost
factor" \citep{Kaiser1987}:

\begin{equation}
f = \left( 1 + \frac{2}{3}\beta + \frac{1}{5}\beta^2 \right),
\label{b:eq:kaiser}
\end{equation}

\noindent where $\beta = \Omega_{\rm m}(z)^{0.55}/b$, with $\Omega_{\rm m}$
denoting the matter density parameter and $b$ the linear bias parameter.  This
expression is overplotted in Fig.~\ref{b:fig:biasz} for comparison. Note that,
in practice, the Kaiser factor is only attained asymptotically
\citep{Jennings2011}, so we again measure the bias in redshift space by
comparing densities in grid cells of size $167\,\Mpc$. Furthermore, there is no
reason to expect this relation to hold for highly nonlinear objects
corresponding to high peaks \citep[e.g.][]{Angulo2005}. 

Despite the scatter among the sub-resolution halo catalogues, we find
reasonably good agreement between the theoretical expectations and the
measurements from the {\tt BASICC} FoF haloes.  Given the comparisons presented
in previous figures, it is not surprising to see the differences for haloes
corresponding to high peaks. Nevertheless, our scheme to assign peculiar
velocities to haloes performs satisfactorily in the regime where the abundance
and clustering in real space are properly imprinted on the sub-resolution
catalogues. This is a remarkable success, extending the usability of the method
to modelling redshift-space distortions.

Now that we have established the range of mass scales over which the sub-resolution halo catalogues give an accurate reproduction of the results seen in high-resolution N-body simulations, in the application of the method presented 
in the next section, we will use a hydrid halo catalogue, made up of 
directly resolved haloes and lower-mass, sub-resolution haloes.

\section{Application: Large scale clustering of Luminous Red Galaxies}
\label{b:sec:lrg}

Recently the clustering of luminous red galaxies (LRG) has been of great
importance in probing different cosmological scenarios. The low number density
but strong clustering of these galaxies means that the spatial distribution of
LRGs can be mapped over vast regions of the sky at relatively low observational
cost. A large survey volume enables  tight constraints to be placed on
cosmological parameters, in particular by measuring the BAO feature
\citep{Eisenstein2005, Gaztanaga2008, Cabre2008, Sanchez09}.  Unfortunately,
there is still an incomplete understanding of the errors associated with
clustering measurements on large scales.  A realistic model for the
uncertainties, including systematic errors,  is crucial to extract cosmological
constraints from the data, since the determination of the best fitting model,
together with the allowed regions in cosmological parameter space, depend
sensitively on the availability of an accurate covariance matrix.

Semi-analytical modelling and observational evidence suggest that LRGs not only
populate very massive haloes but they can also be found in haloes with masses
as small as $10^{11}\,\Mass$ albeit with a low probability \citep{Almeida2008,
Wake2008}. Therefore, the modelling of LRG clustering, and the BAO feature
imprinted on it, requires huge simulations with a considerable dynamic range in
mass.  Although such extremes can be achieved in modern supercomputers these
tend to be one-off runs and the computational cost is enhanced to inaccessible
levels when studying uncertainties or subtle features present in the clustering
which require many realizations. 

In this section we approach this problem using the algorithm we described
above. Specifically, we generate $50$ LRG catalogues to compute the mean and
variance of the two-point correlation function.  Details of the creation of the
LRG catalogues as well as the clustering measurements are presented in the
following subsections.

\subsection{The haloes and LRG catalogues} \label{b:sec:lrg:catalogue}

\begin{figure} 
\includegraphics[width=8.5cm]{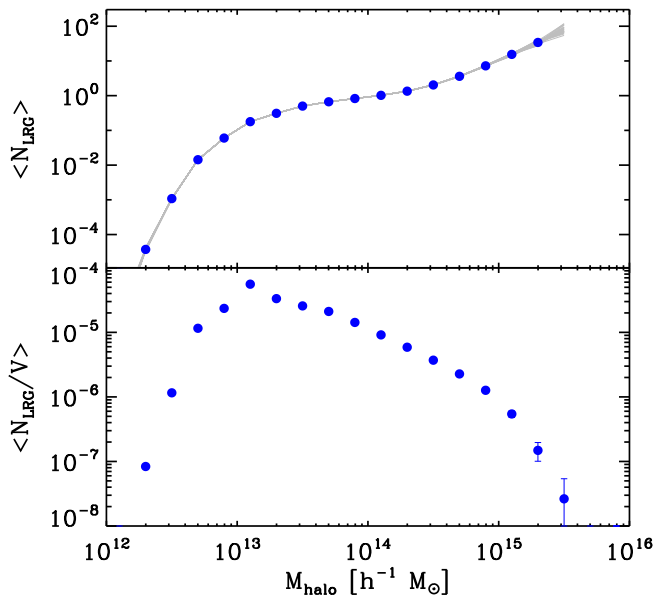}
\caption[Number of LRGs as a function of the host halo mass.]{The application 
of the method to the construction of LRG catalogues. Here we use a hybrid
catalogue comprised of sub-resolution haloes and haloes which are directly
resolved in the simulations. Top: the mean number of LRGs per halo as a
function of the host halo mass in our simulations.  Bottom: the total number
density of LRGs as a function of the host halo mass. The grey lines in the top
panel show the quantities in each of our simulations while the blue symbols
show the mean and dispersion.  \label{b:fig:lrg_hod}}
\end{figure}

The starting point in the creation of the LRG mock catalogues is to predict the
abundance and spatial distribution of the DM haloes that are likely to host
such galaxies. For this purpose we created $50$ hybrid halo  catalogues, each
one spanning $4$ orders of magnitude in mass within a volume of
$2.4\,h^{-3} {\rm Gpc}^{3}$ at $z=0.5$.

The halo catalogues are hybrid in the sense that they consist of two types of
haloes. The high mass ones ($M > 1.85 \times 10^{13}\,\Mass$  correspond to
objects identified directly using a FoF algorithm, with at least $10$
particles, in each of the {\rm L-BASICC} simulations. Then, smaller masses
sub-resolution haloes ($5.48\times10^{11} < M/(\Mass) < 1.85 \times 10^{13}$)
were created using the algorithm described in \S\ref{b:sec:method}. We
recall that our method is accurate to better than the $10\%$ level 
for this mass range.
In this way, we are effectively extending the dynamic range of the {\rm
L-BASICC} simulations towards lower masses. Combining the two types of haloes
also eliminates the need to reproduce high mass haloes in the sub-resolution
catalogues, which proved to be troublesome (see Section~3.2).

Once we have generated the catalogues that contain all the haloes that are
expected to host LRGs, we use a Halo Occupation Distribution (HOD) model to
determine how many LRGs on average populate each DM halo \citep[for a review of
the halo model see ][]{Cooray2002}.  Following \cite{Wake2008} we can express
the mean number of central LRGs, $N_{\rm c}$ as a function of the host halo
mass, $M_{\rm halo}$ as:

\begin{equation}
\langle N_{\rm c} \vert M_{\rm halo} \rangle = \exp( - M_{\rm min}/ M_{\rm halo} ),
\label{b:eq:lrg:nc}
\end{equation}

\noindent and the mean number of satellite LRGs, $N_{\rm s}$ as:

\begin{equation}
\langle N_{\rm s} \vert M_{\rm halo} \rangle = ( M_{\rm halo}/M_1)^{\alpha}.
\label{b:eq:lrg:ns}
\end{equation}

\noindent Consequently, the total number of LRGs has an expected value of 

\begin{equation}
\langle N_{\rm LRG} \vert M_{\rm halo} \rangle = \langle N_{\rm c} \vert M_{\rm halo} \rangle [ 1 + \langle N_{\rm s} \vert M_{\rm halo} \rangle ], 
\label{b:eq:lrg:nt}
\end{equation}

\noindent where $\alpha$, $M_{\rm min}$ and $M_1$ are, in principle, free
parameters that can be constrained either by comparing to observational
estimates of clustering observations or through semi-analytical galaxy
formation modelling.  Indeed, \cite{Wake2008}, using the measured clustering of
2SLAQ LRGs \citep{Cannon2006}, found that the  best fitting values of $M_{\rm
min}$, $M_1$ and $\alpha$ are $2.19 \times 10^{13}\,\Mass$, $2.82 \times
10^{13}\,\Mass$ and $1.86$ respectively. In the second step, we assume that
Eq.~\ref{b:eq:lrg:nc} and \ref{b:eq:lrg:ns} follow a Poisson distribution,
which, combined with the values from Wake et al., allows us to place LRGs
within our hybrid halo catalogues. Note that the alternative approach of
applying full semi-analytic modelling to the hybrid halo catalogues could
also have been taken.

Each of our final catalogues contain $398 \, 963$ galaxies, or equivalently, a
number density of $1.66\times10^{-4} h^{3}\,{\rm Mpc^{-3}}$. Even though, on
average, there is less than one LRG per sub-resolution halo, together the
sub-resolution haloes host a total of $114\,243$ galaxies which represents
$28\%$ of whole LRG sample.  Fig.~\ref{b:fig:lrg_hod} shows the resulting mean
number of LRG in our catalogues per halo (top panel) as well as the total
number LRGs (bottom panel), in both cases as a function of the mass of the host
halo. 

\subsubsection{Correlation Function of LRGs}

\begin{figure} 
\includegraphics[width=8.5cm]{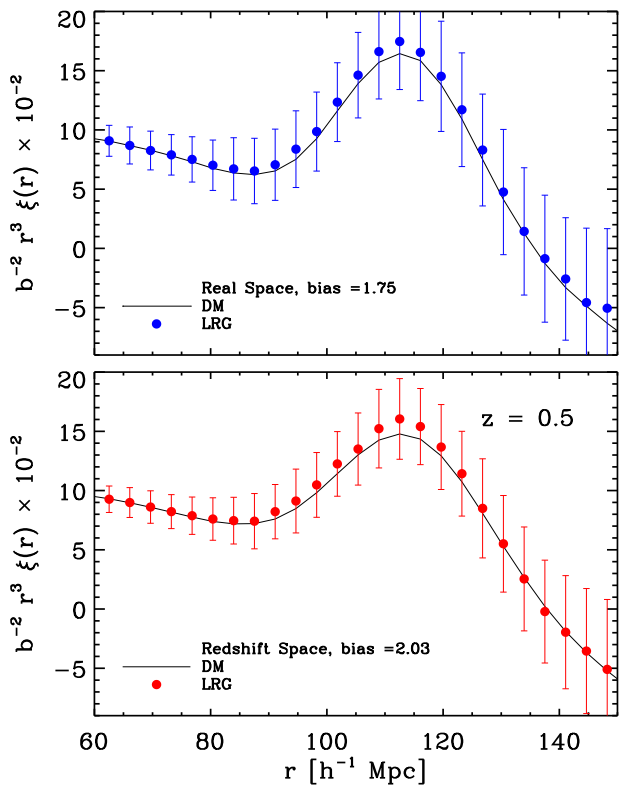}
\caption[The mean and variance of the LRG correlation function from our
catalogues at $z=0.5$.]{ The mean and variance of the correlation function
measured from LRG samples constructed from our hybrid catalogues of 
resolved and sub-resolution haloes at $z=0.5$. The top panel shows the 
real space correlation function and the bottom panel shows redshift space. 
In each case filled circles indicate the clustering measured from 
the LRG catalogues created by populating hybrid halo catalogues 
(i.e. the mixture of sub-resolution and FoF haloes) with the LRG HOD inferred
by \cite{Wake2008}. The solid lines show the mean correlation function
measured from the dark matter particles of all 50 {\tt L-BASICC} realisations.
To allow a full comparison, we have divided each measurement (and the
respective variance) by a constant bias, measured in the range $r =
[60-70]\,h^{-1}$ Mpc, and by the expected Kaiser boost factor in the
case of redshift-space measurements}. Note that we display $\xi(r) \times r^{3}$ in the y-axis
to enhance the BAO peak.
\label{b:fig:lrg_corr} 
\end{figure}

At this point, we are now in a position to investigate the clustering of
LRGs.  We measure the correlation function  using Fast Fourier Transforms.
This approach is considerably more efficient than computations carried out in
configuration space, when one is interested in the correlation function 
on large scales measured from catalogues containing a large number of objects. 

In brief, the method uses a pixelization of the density field from which the
spherically averaged correlation function can be estimated from the amplitude
of Fourier modes as: 

\begin{equation}
\xi(\vect{r}) = \mathcal{F}^{-1}\left\{||\mathcal{F}[\delta(\vect{x})]||\right\}
\end{equation}

\noindent where $\delta = (n(x) - \langle n \rangle) / \langle n \rangle$ is
the density fluctuation on a grid, and $\mathcal{F}[\delta]$ is its Fourier
transform. Vertical bars denote the modulus of a complex field, and
$\mathcal{F}^{-1}$ an inverse Fourier Transform. We carry out this operation 
using a Fast Fourier transform with a grid of dimensions $N_{\rm grid} = 1024$,
which corresponds to $1.3 h^{-1}$Mpc for the {\tt L-BASICC} simulation box
size.  This method gives an accurate estimation of the correlation function for
scales larger than a few grid cells. 

Fig.~\ref{b:fig:lrg_corr} shows the result of applying this procedure to
compute the correlation function for LRGs in each of our $50$ catalogues. The
top panel displays the measurements in real space while the bottom panel shows
redshift space. In both cases the mean and variance of the measurements are
indicated by the filled circles and error bars. In order to assess our results,
we have measured the correlation function for subsets of DM particles at
$z=0.5$ from the {\tt L-BASICC} simulations. We display the mean of all
$50$ simulations in real and redshift space as a solid line in the top and
bottom panels, respectively. This allows us to compare the form of the
correlation function measured from our LRG catalogues with that of the
underlying dark matter distribution. Note that the y-axis shows $\xi \times
r^{3}$ instead of $\xi$, as in this way the acoustic peak is highlighted. In
addition, the results (including the errors) in both real and redshift space
have been renormalized as described in the figure caption. 

\begin{figure} 
\includegraphics[width=8.5cm]{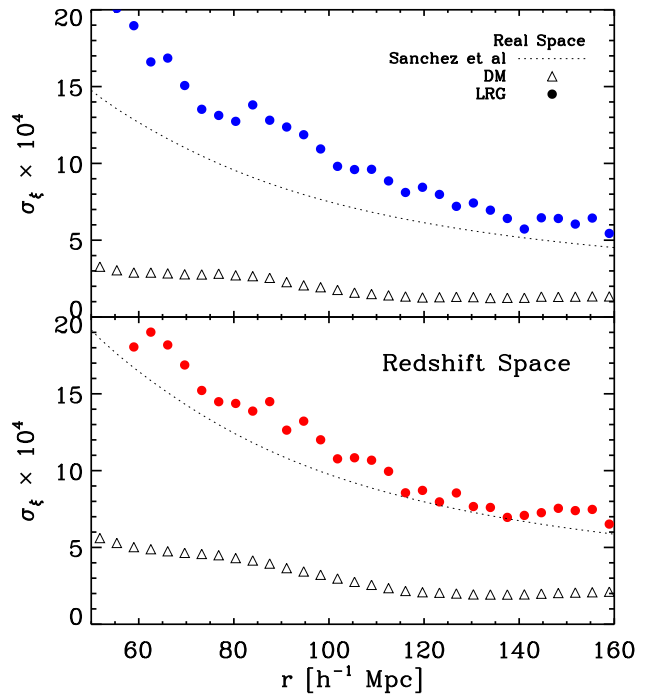}
\caption{The variance in the 2-pt
correlation function measured from $50$ LRGs catalogues in real space (top) and
redshift space (bottom) at $z=0.5$, constructed using the hybrid catalogues of
resolved and sub-resolution haloes. We also plot a theoretical estimate for
the variance from Sanchez et al. (2008; dotted lines) 
\label{b:fig:lrg_var}}
\end{figure}

By comparing the correlation function of LRGs with that of the dark matter we
can see the effects of galaxy bias.  Fig.~\ref{b:fig:lrg_corr} shows that the
respective correlation functions, after applying a scaling in amplitude, agree
fairly well with one another, implying that the LRG bias is approximately scale
independent over the range of pair separations plotted.  There is a small
residual dependence of the bias on scale in real space which seems to be
accentuated in redshift space. Although the discrepancy is not significant
given the size of the errors associated with the simulation volume, using a
simulation with $10$ times larger volume and $3375$ times more particles,
(Angulo et al 2013, in prep.) recently showed that distortions of this type are
expected in biased tracers of the DM field \citep[see
also][]{Padmanabhan2009,Mehta2011}. This scale-dependent bias, absent in 
approaches that simply apply a biasing scheme on top of the DM field, is an
example of the benefits of an hybrid approach like the one proposed here.

\begin{figure*} 
\includegraphics[width=16.5cm]{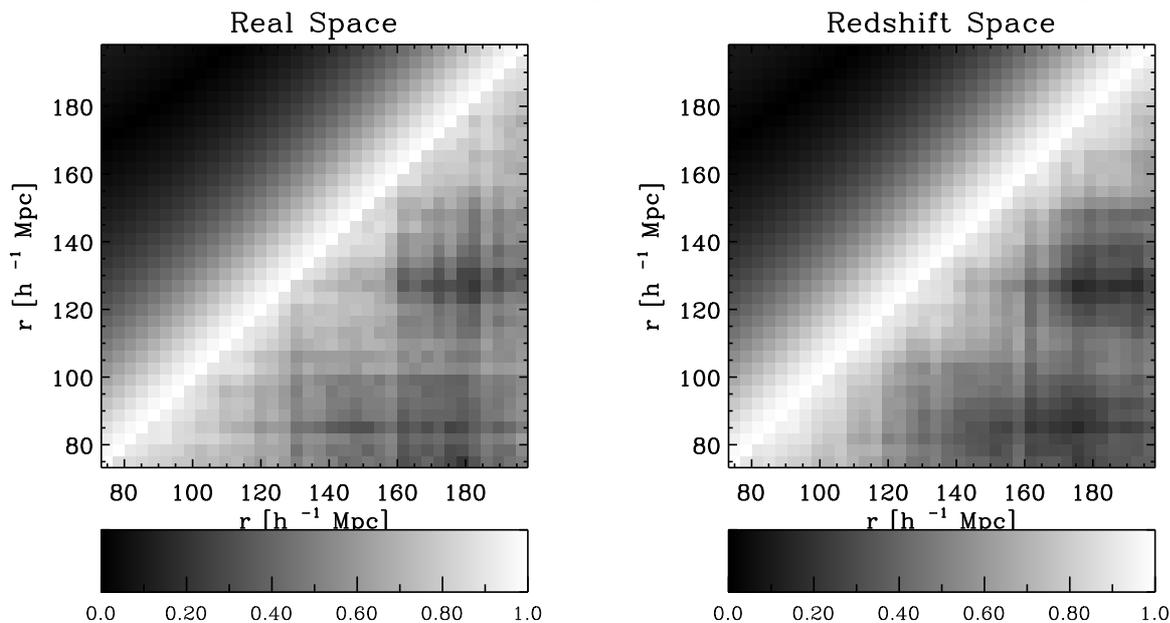}
\caption{The normalized covariance matrix from our ensemble of LRG mock
catalogues, constructed from the hybrid catalogues of resolved and
sub-resolution haloes (lower triangular region) and from a analytical
prediction from Snachez et al (2008) that incorporates the correct volume,
bias, number density and binning (upper triangular region).  The left plot show
displays the results in real space while the left displays the covariance
matrix in redshift space. 
\label{b:fig:lrg_cov}}
\end{figure*}

In Fig.~\ref{b:fig:lrg_var} we compare the variance measured from our
ensemble of LRG catalogues (filled circles) with that measured from the dark
matter samples (triangles). By comparing both measurement we illustrate the
importance of shot-noise in the expected variance. The dotted line shows a
theoretical prediction for the variance based on power spectrum measurements
which include the effects of finite number of modes, discrete noise, bias and
binning \citep[see ][for more details]{Sanchez2008}. The theoretical
predictions by \cite{Sanchez2008} provide a fairly good match to the variance
in our LRG samples, showing that our catalogues have the expected variance. 

We extend this comparison in Fig.~\ref{b:fig:lrg_cov} in which we display the
normalized covariance matrix \citep{Cohn2006, Smith2008}, $C_{\xi}(r,r') \equiv
\langle (\xi(r) - \bar{\xi(r)}) (\xi(r') - \bar{\xi(r')})
\rangle/\sigma(r)/\sigma(r')$, in real space (left plot) and in redshift space
(right plot). The above diagonal part of the plot shows  the expected
covariance as computed following \cite{Sanchez2008}. The below diagonal part
shows the covariance for the LRG catalogues. The non-diagonal parts of the
covariance matrix show good agreement between the LRG and the theoretical
expectations, similar to the case with the comparison of the variance. However,
our LRG catalogues show stronger off-diagonal correlations than the expectation
and also show more structure, in particular an excess correlation at the BAO
location. One possible explanation could be the contribution of the higher
order moments of the halo density field, which are present in our LRG samples 
but absent in the Sanchez et al. predictions. As shown by \cite{Angulo2008b},
the higher order moments of haloes differ considerably from those of DM. As an
example, recall that even if the DM  density field is Gaussian (i.e. the higher
moments are zero), then haloes will have non-zero higher order correlations
which contribute to the covariance matrix. Nevertheless, the results are still
noisy given the small number of simulations in our ensemble and further
investigation is required. In any case, the performance of our catalogues is
remarkable and illustrates the feasibility of constructing detailed covariance
matrices from computationally-cheap $N$-body simulations that have the correct
diagonal terms.

\section{Summary} \label{b:sec:conc}

Due to the large volumes that future surveys are expected to map, the resulting
measurements of galaxy clustering will be of exquisite accuracy, with the
target of distinguishing between different models for the acceleration of the
cosmic expansion. The clustering signals predicted by competing models often
differ by small amounts. It is therefore essential to understand the systematic
and sampling errors associated with the measurements. Only in this way will it
be possible to extract robust conclusions from the data. In practice, this
challenge can only be met by techniques which make use of cosmological $N$-body
simulations, since this approach gives the best estimate of the contribution of
various nonlinear effects to the measured clustering. 

We have devised and illustrated the feasibility of a scheme that allows the
rapid and efficient creation of large numbers of galaxy mock catalogues which
are able to resolve all of the galaxies selected in upcoming surveys. This is
done by taking moderate resolution simulations and effectively extending their
dynamic range in halo mass to mimic running a simulation with a substantially
larger number of particles. Our method uses the density field extracted from
the moderate resolution $N$-body simulation and combines it with the bias
parameters and mass functions extracted from a higher resolution simulation. In
this way, it is possible to predict statistically the expected density field of
dark matter haloes in the moderate resolution simulation volume. Since
low-resolution simulations are relatively easy to generate, our procedure
allows the investigation of uncertainties in both the measurements themselves
and in the procedures employed to extract robust information from the data. 

We have shown that, on large scales, the generated halo population agrees with
the population seen directly in a high resolution simulation over a
considerable range of masses.  At $z=0$ in particular, the abundance and
clustering strength, in both real and redshift space, of haloes less massive
than $7.51\times10^{13}\,\Mass$ agree to within $10$\% with those computed
directly from FoF haloes identified in a high-resolution simulation. For high
mass haloes or at higher redshifts, our procedure performs less satisfactorily.

An interesting application of our scheme is to the creation of hybrid halo
catalogues. High mass haloes can be extracted directly from cosmological
$N$-body simulations, whilst low mass haloes which lie beyond the grasp of the
simulation can be generated using our technique.  In this way, we
can employ our algorithm in the regime where it works best.  As an example, we
have created $50$ such catalogues from the {\tt L-BASICC} simulations which are
combined with a HOD for LRGs.  From the resulting galaxy catalogues we are
successfully able to predict their mean correlation function along with the
full covariance matrix. We found that the variance in a sample of dark matter
particles drawn from the simulations and analytical estimates are in agreement
with measurements from the LRG catalogues.  In spite of this, differences in
the off-diagonal terms of the covariance matrix were found.

In the LRG example presented, we extended the halo mass resolution of the {\tt
L-BASICC} runs by a factor of $\approx 30$, since this was all that was
demanded by this application. The specifications of the available high
resolution $N$-body simulation set the limit on the boost attainable in the
resolution of the moderate resolution runs. For example, if we had instead
chosen to augment the {\tt L-BASICC} runs using the Millennium-II simulation of
\cite{Boylan-Kolchin2009}, which modelled the growth of structure using
$2160^3$ particles in a volume of $(100)^{3} h^{-3} {\rm Mpc}^{3}$, then the
resulting halo catalogue would be the equivalent of that expected from a
simulation employing $28944^{3}$ or more than 24 trillion particles. This is
around 50 times larger than the largest number of particles used in an $N$-body
simulation to date. Our approach will allow the production of halo catalogues
equivalent to running large numbers of such simulations. 

The algorithm presented here is already useful for generating mock
observations and in creating covariance matrices, particularly if combined 
with novel techniques to mimic running very large ensembles of 
simulations \citep[e.g.][]{Schneider11}. Nevertheless, it 
could be enhanced in a number of ways, including the following:

\begin{itemize}
\item The placement of haloes within the smoothing volume could be
improved by distributing halos following a given correlation function.
\item Halos could be placed recursively using different smoothing
scales, starting with the whole box and stoping at any desired scale. 
Here, a parent cell puts contraints on all their child cells, which 
could be used to include an arbitrary scale-dependent biasing scheme.
\item A more complex biasing scheme could be implemented, which can
be calibrated directly using $N$-body simulations, and could be different
for halos of different mass. Dependences in addition to the density, 
such as the tidal field could be taken into account.
\item The form of the probability distribution function of halos 
given a DM overdensity can be calibrated directly with $N$-body 
simulations instead of making the assumption that this has a 
standard form such as a Poisson distribution. 
\item Similarly, the covariance matrix among different halo mass bins can
be used in sampling the DM-halo relationship.
\item Extended features can be incorporated such as exclusion effects between haloes, 
an additional density-dependent velocity dispersion, and the substructure
content of halos.
\end{itemize}

Nevertheless, even without these improvements, we expect that the simple
technique presented in this paper will improve the understanding and treatment
of uncertainties in observations and, therefore, will allow the full potential
of measurements of the large scale distribution of galaxies to be reached.

\section*{Acknowledgments}
The calculations for this paper were performed on the ICC Cosmology
Machine, which is part of the DiRAC Facility jointly
funded by STFC, the Large Facilities Capital Fund of BIS,
and Durham University. We acknowledge support from the
Durham STFC rolling grant in theoretical cosmology.

\bibliographystyle{mn2e} \bibliography{database}

\label{lastpage} \end{document}